\documentclass[10pt,letterpaper,english,accepted=2021-11-24]{quantumarticle}

\makeatletter

\pdfoutput=1
\usepackage{braket} 
\usepackage[numbers,sort&compress]{natbib}
\newcommand{\up}{\uparrow}
\newcommand{\dw}{\downarrow}

\usepackage[utf8]{inputenc}
\usepackage[english]{babel}
\usepackage[T1]{fontenc}
\usepackage{amsmath}

\usepackage{units}
\usepackage{amsmath}
\usepackage{amssymb}
\usepackage{upgreek}

\makeatother

\begin{document}
\title{Dynamical second-order noise sweetspots in resonantly driven spin
qubits}

\author{Jordi Picó-Cortés}
\affiliation{Instituto de Ciencia de Materiales de Madrid (CSIC) 28049, Madrid, Spain.}
\affiliation{Institute for Theoretical Physics, University of Regensburg, 93040
	Regensburg, Germany.}
\orcid{0000-0002-7705-3966}
\email{Jordi.Pico-Cortes@physik.uni-regensburg.de}
\author{Gloria Platero}
\affiliation{Instituto de Ciencia de Materiales de Madrid (CSIC) 28049, Madrid, Spain.}

\maketitle

\begin{abstract}
	Quantum dot-based quantum computation employs extensively the exchange
	interaction between nearby electronic spins in order to manipulate
	and couple different qubits. The exchange interaction, however, couples
	the qubit states to charge noise, which reduces the fidelity of the
	quantum gates that employ it. The effect of charge noise can be mitigated
	by working at noise sweetspots in which the sensitivity to charge
	variations is reduced. In this work we study the response to charge
	noise of a double quantum dot based qubit in the presence of ac gates,
	with arbitrary driving amplitudes, applied either to the dot levels
	or to the tunneling barrier. Tuning with an ac driving allows to manipulate
	the sign and strength of the exchange interaction as well as its coupling
	to environmental electric noise. Moreover, we show the possibility
	of inducing a second-order sweetspot in the resonant spin-triplet
	qubit in which the dephasing time is significantly increased. 
\end{abstract}

\section{Introduction}

\begin{figure}[t]
	\begin{centering}
		\includegraphics[width=1\columnwidth]{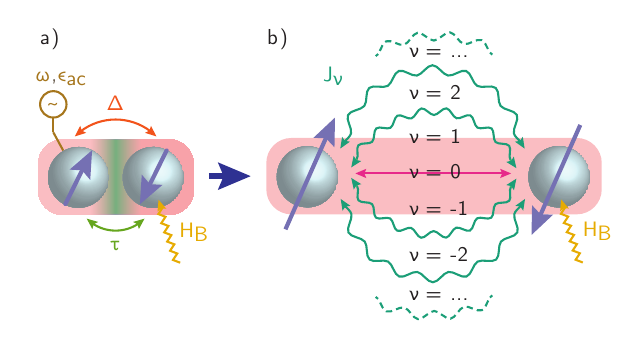}
		\par\end{centering}
	\caption{a) Scheme of the system. A DQD with a single electron spin defined
		in each dot and a tunneling amplitude $\tau$ between both is set
		under a magnetic field gradient $\Delta$. The leftmost dot is subjected
		to an ac gate of amplitude $\epsilon_{\text{ac}}$ and frequency $\omega$.
		The system is coupled to a bosonic bath $\hat{H}_{B}$, denoted by
		a yellow arrow, accounting for the effect of charge noise. b) Scheme
		of the exchange interaction with an ac field included. The different
		sidebands (marked in green) correspond to processes involving the
		absorption or emission of a photon and together contribute to the
		total exchange rate. \label{fig:DQD-Scheme}}
\end{figure}

Quantum computation in quantum dot arrays makes extensive use of the
exchange interaction between the electron spins both for one\citep{DiVincenzo2000,Levy2002}
and two-qubit gates\citep{Lui2012,Wardrop2014}. Unfortunately, the
exchange interaction couples the qubit states to electric noise due
to fluctuations in the charging energy\citep{Stopa1998,Hu2006}, defects
in the dot\citep{Hayashi2003,Culcer2013} and the interaction with
the electric environment\citep{Yurkevich2010} resulting in reduced
coherence times\citep{Dial2013,Qi2017}. Moreover, the exchange interaction
may result in double occupancy \citep{Schliemann2001,Barrett2002}
and leakage errors\citep{Rebentrost2009,Chasseur2015,Mehl2015}. Nonetheless,
it has been employed to reach large gate fidelities in quantum dot-based
qubits\citep{Nichol2017,Huang2019} and forms the basis of an extending
number of proposals for solid state quantum computation\citep{Kloeffel2013,Shim2016,Russ2017,Sala2017,Pan2020,Sala2020}.

One promising venue for improvement is the use of ac gates in order
to tune the properties of the quantum dot array\citep{GomezLeon2020,Qiao2021},
in what has been termed Floquet engineering\citep{Benito2014,Eckardt2015,Oka2019,PerezGonzalez2019}
of quantum systems. In the presence of ac gates, electron tunneling
is accompanied by the absorption or emission of photons, resulting
in a new set of photoassisted paths, called sidebands\citep{Platero2004,GallegoMarcos2015}.
The exchange interaction, as a virtual tunneling process, is similarly
modified by the ac driving\citep{Sanchez2014B,Stano2015}. Several
qubit platforms employing a resonant exchange interaction have been
studied both theoretically and experimentally\citep{Klauser2006,Kim2015,Song2016,Zajac2016}.
Resonant exchange qubits based on double quantum dots in both GaAs/AlGaAs
heterostructures\citep{Shulman2014} and Silicon\citep{Takeda2020},
as well as on triple quantum dots\citep{Doherty2013,Taylor2013} have
been developed. The latter allows for qubit operation involving only
the exchange interaction and show excellent protection against electric
noise\citep{Russ2015}. Similarly, resonant exchange CNOT gates have
been developed and implemented experimentally which employ the same
method of operation\citep{Zajac2017,Russ2018b}. 

In the recent years, several works\citep{Jing2014,Yang2017,Frees2019,Mundada2020,Huang2021}
have developed the idea of employing ac gates to engineer noise sweetspots\citep{Makhlin2004,Fei2015}
in which the decoherence time due to noise is significantly increased.
A recent paper\citep{Pico-Cortes2019} shows how an ac voltage can
induce a dynamical sweetspot due to sideband interference, suppressing
the coupling to the electric environment. In this work, we study how
to employ ac gates to induce dynamical sweetspots in a qubit defined
on a double quantum dot (DQD). The DQD qubit is an ideal platform
to analyze dynamical sweetspots in the photoassisted exchange interaction
both for its simplicity and because electric noise is often the most
important source of decoherence, in particular for purified silicon\citep{Tyryshkin2011,Veldhorst2014}.
The DQD-based resonant qubit is operated by tuning the frequency of
the oscillating exchange interaction in resonance with the splitting
between the $\ket{\up,\dw},\ket{\dw,\up}$ states due to a magnetic
field gradient. We consider the effect of modulating both the dot
gates and the tunneling barriers with an ac voltage, and investigate
resonances involving an arbitrary number of photons. Moreover, we
show how an ac gate can be used to engineer the coupling between the
electric environment and the DQD. In particular, when driven with
a particular amplitude of the ac gate, we find that the DQD exhibits
a second-order sweetspot\citep{2104.07485} in which the qubit is
unaffected by noise up to second order in the coupling with the electric
environment, significantly increasing the dephasing time of the qubit. 

The paper is organized as follows. In Sec.~\ref{sec:Theoretical-Model}
we introduce the theoretical model that we will be employing. In Sec.~\ref{subsec:Resonant-operation}
the basic features of the qubit operation are introduced. In particular,
we consider the case of a resonant transition between the two $S_{z}=0$
states with one unpaired spin in each dot mediated by the ac voltage.
In Sec.~\ref{subsec:Operation-under-charge} we explore the different
ways in which the ac gate can be used to improve the qubit operation
under charge noise for the cases where the ac gate voltages are applied
to the individual quantum dots and to the tunneling gates between
them, respectively. Finally, in Sec.~\ref{sec:Discussion} we summarize
the results and discuss the experimental implementation of the proposal. 

\section{Theoretical Model\label{sec:Theoretical-Model}}

We consider the Hamiltonian for a DQD in an extended Hubbard model
with nearest neighbor tunneling $\tau$ between the two sites, denoted
by $\alpha=1,2$, Zeeman splittings $E_{z,\alpha}$, and both intra-dot
and inter-dot interactions, $U_{\alpha}$ and $V$ respectively. The
dot in the left is subjected to an ac voltage of amplitude $\epsilon_{\text{ac}}$,
frequency $\omega$, and with an initial phase $\phi$. The Hamiltonian
governing the DQD dynamics is then given by
\begin{align}
	\hat{H}\left(t\right) & =\hat{H}_{0}+\hat{H}_{\mathrm{ac}}\left(t\right)+\hat{H}_{U}+\hat{H}_{T},\label{eq:Ht}\\
	\hat{H}_{0} & =\sum_{\alpha;\sigma}\epsilon_{\alpha}\hat{n}_{\alpha,\sigma}+\sum_{\alpha}\hbar^{-1}E_{z,\alpha}\hat{S}_{z,\alpha},\nonumber \\
	\hat{H}_{U} & =\sum_{\alpha}U_{\alpha}\hat{n}_{\alpha,\uparrow}\hat{n}_{\alpha,\downarrow}+\sum_{\alpha<\beta;\sigma,\sigma'}V\hat{n}_{\alpha,\sigma}\hat{n}_{\beta,\sigma'}\nonumber \\
	\hat{H}_{T} & =\sum_{\alpha,\beta;\sigma}\tau\left(\hat{c}_{\alpha,\sigma}^{\dagger}\hat{c}_{\beta,\sigma}+\mathrm{H.c.}\right),\nonumber 
\end{align}
where $\hat{c}_{\alpha,\sigma}^{\dagger}\left(\hat{c}_{\alpha,\sigma}\right)$
is the fermion creation (annihilation) operator for an electron in
dot $\alpha$ with spin $\sigma=\up,\dw$ and $\hat{n}_{\alpha,\sigma}=\hat{c}_{\alpha,\sigma}^{\dagger}\hat{c}_{\alpha,\sigma}$.
The ac gates can be introduced either as a quantum dot gate $\hat{H}_{\mathrm{ac}}\left(t\right)=\sum_{\sigma}\epsilon_{\text{ac}}\hat{n}_{1,\sigma}\cos\left(\omega t+\phi\right)$
or in the tunneling amplitude as\footnote{The dependence of the tunneling amplitude on the gate voltage may
	also be non-trivial\citep{Zajac2017}, resulting in higher harmonics
	in $\tau\left(t\right)$ and the possibility of resonances involving
	a higher number of photons.} $\hat{H}_{\mathrm{ac}}\left(t\right)=(\tau_{\text{ac}}/\tau)\cos\left(\omega t+\phi\right)\hat{H}_{T}$.
In actual experimental conditions, both are expected to be present\citep{Zhang2018}
as they appear as a result of modulations of the same interconnected
potential. We will consider them separately here for reasons of simplicity.
The DQD with all processes in Eq.~\ref{eq:Ht} is represented schematically
in Fig.~\ref{fig:DQD-Scheme}~a). 

Considering that the DQD can be occupied by a maximum of two electrons,
the relevant states have either spin projection $\left|S_{z}\right|=1$,
$\left\{ \ket{\up,\up},\ket{\dw,\dw}\right\} $ or $S_{z}=0$, including
the two qubit states $\mathcal{Q}=\left\{ \ket{\up,\dw},\ket{\dw,\up}\right\} $
and  two states with double occupancy, $\mathcal{D}=\left\{ \ket{\up\dw,0},\ket{0,\up\dw}\right\} $,
where $\ket{\sigma,\sigma'}$ is the state with spin $\sigma$ in
the left dot and spin $\sigma'$ in the right dot. The two states
with $\left|S_{z}\right|=1$, are not connected to the rest by any
term in the ideal Hamiltonian of Eq.~\ref{eq:Ht}. In realistic conditions,
the $S_{z}=0$ and the $|S_{z}|=1$ subspaces are connected through
coupling to the environment, including the nuclear spin bath coupled
through the hyperfine interaction and phonons coupled through the
spin-orbit interaction or magnetic field gradients. We will consider
that the qubit is operated in a time scale shorter than the relaxation,
leakage and dephasing times associated with these processes ---which
can be quite large for isotopically purified Silicon\citep{Tyryshkin2011,Veldhorst2014}
or if spin-echo sequences are used\citep{Bluhm2010,deLange2010}---
and focus on the $S_{z}=0$ subspace. However, Silicon carries its
own complications in the form of valley physics\citep{Culcer2009,Li2010}.
The valley splitting in Silicon can nonetheless be tuned by the use
of gate potentials resulting in high inter-valley relaxation lifetimes\citep{Goswami2006,Yang2013}. 

\section{Results\label{sec:Results}}

\subsection{Qubit operation with ac gates\label{subsec:Resonant-operation}}

The qubit, consisting of the states $\ket{\up,\dw}$ and $\ket{\dw,\up}$,
is initialized in the ground state of the static qubit. The ac bias
is then tuned to the resonance frequency between the qubit states.
Initialization can be performed adiabatically in order to ensure that
the dressed states are mapped correctly from to the static eigenstates\citep{Deng2015,Mundada2020,2108.11260}.
Similarly, readout can be performed by adiabatically turning off the
ac voltage. The physics in the resonance can be studied in a convenient
time-independent framework by employing an effective cotunneling Hamiltonian
in the Rotating Wave Approximation (RWA). We consider that the ac
voltage is applied in the left quantum dot in a resonance involving
$n$ photons, but allow for a small offset of the resonance (compared
to the ac voltage frequency). Then, the qubit Hamiltonian can be written
as (See Appendix~\ref{sec:Effective-cotunneling-Hamiltonia})
\begin{align}
	\hat{H}_{\mathcal{Q}}^{\left(0\right)} & =\left(\Delta E-n\hbar\omega\right)\hat{\sigma}_{z}\nonumber \\
	& +\mathcal{J}_{n}\cos\left(n\phi\right)\hat{\sigma}_{x}+\mathcal{J}_{n}\sin\left(n\phi\right)\hat{\sigma}_{y},\label{eq:HQ}
\end{align}
where $\hat{\sigma}_{k}$, $k=x,y,z$ are the Pauli matrices in the
basis $\left\{ \ket{\up,\dw},\ket{\dw,\up}\right\} $ and 
\begin{figure}
	\begin{centering}
		\includegraphics[width=1\columnwidth]{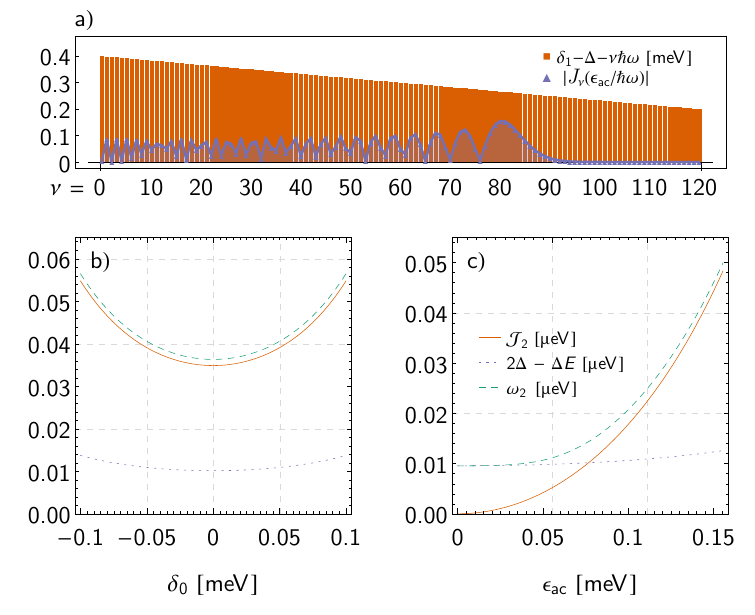}
		\par\end{centering}
	\caption{a) In red bars, the energy difference between $\ket{\up,\dw}$ and
		$\ket{\up\dw,0}$ (i.e., the lowest contribution to the denominators
		in Eqs.~\ref{eq:ERWA} and~\ref{eq:JRWA}) for each sideband transition
		$\nu$ at $\delta_{0}=0$. In blue lines, the absolute value of the
		corresponding Bessel function $J_{\nu}\left(\epsilon_{\text{ac}}/\hbar\omega\right)$
		for $\epsilon_{\text{ac}}=0.14$ meV at the $n=2$ resonance (see
		text). For $\nu\simeq120$ the Bessel functions are negligible, far
		from the photoassisted resonance, indicating that the system is properly
		described including up to this value. b) The exchange interaction
		$\mathcal{J}_{2}$ (red, continuous), the correction to the energy
		splitting $\Delta-\Delta E$ (blue, dotted), and the qubit energy
		$\omega_{2}=\sqrt{\mathcal{J}_{2}^{2}+\left(\Delta E-2\hbar\omega\right)^{2}}$
		(green, dashed, see text) as a function of $\delta_{0}$ at $\epsilon_{\text{ac}}=0.14\,\text{meV}$.
		c) The same three quantities as a function of the ac voltage amplitude
		$\epsilon_{\text{ac}}$ at $\delta_{0}=0.085\,\text{meV}$. Parameters
		employed: $\Delta=1.66\bar{6}$ $\mu$eV, $U_{1}=U_{2}=0.5$ meV,
		$V=0.1$ meV, $\tau=0.01~\text{meV}$, and~$\hbar\omega=\Delta$.
		\label{fig:0th-order}}
\end{figure}

\begin{align}
	\Delta E & =E^{+}-E^{-},\\
	E^{\pm} & =\pm\Delta\nonumber \\
	& -2\tau^{2}\sum_{\nu=-\infty}^{\infty}\left[\frac{J_{\nu}^{2}\left(\frac{\epsilon_{\text{ac}}}{\hbar\omega}\right)}{\delta_{2}\mp\Delta-\nu\hbar\omega}+\frac{J_{\nu}^{2}\left(\frac{\epsilon_{\text{ac}}}{\hbar\omega}\right)}{\delta_{1}\mp\Delta+\nu\hbar\omega}\right],\label{eq:ERWA}
\end{align}
\begin{align}
	\mathcal{J}_{n}= & \tau^{2}\sum_{\nu=-\infty}^{\infty}\nonumber \\
	\times & \left[\frac{J_{\nu}\left(\frac{\epsilon_{\text{ac}}}{\hbar\omega}\right)J_{\nu-n}\left(\frac{\epsilon_{\text{ac}}}{\hbar\omega}\right)}{\delta_{1}-\Delta+\nu\hbar\omega}+\frac{J_{\nu}\left(\frac{\epsilon_{\text{ac}}}{\hbar\omega}\right)J_{\nu-n}\left(\frac{\epsilon_{\text{ac}}}{\hbar\omega}\right)}{\delta_{2}+\Delta-\nu\hbar\omega}\right.\nonumber \\
	+ & \left.\frac{J_{\nu}\left(\frac{\epsilon_{\text{ac}}}{\hbar\omega}\right)J_{\nu+n}\left(\frac{\epsilon_{\text{ac}}}{\hbar\omega}\right)}{\delta_{2}-\Delta-\nu\hbar\omega}+\frac{J_{\nu}\left(\frac{\epsilon_{\text{ac}}}{\hbar\omega}\right)J_{\nu+n}\left(\frac{\epsilon_{\text{ac}}}{\hbar\omega}\right)}{\delta_{1}+\Delta+\nu\hbar\omega}\right].\label{eq:JRWA}
\end{align}
Here, $J_{\nu}\left(z\right)$ is the $\nu$th  Bessel function of
the first kind, $\Delta=E_{z,1}-E_{z,2}$, $\delta_{0}=\epsilon_{2}-\epsilon_{1}$,
$\delta_{1}=U_{1}-V-\delta_{0}$ and $\delta_{2}=U_{2}-V+\delta_{0}$,
corresponding to the gradient splitting, the detuning between the
dot levels and the energy difference between the $S_{z}=0$ states
with $\left(1,1\right)$ occupation and the ones with $\left(2,0\right)$
and $\left(0,2\right)$, respectively. The qubit defined in Eq.~\ref{eq:HQ}
can be controlled by a proper alignment of the ac field phase $\phi$.
Alternatively, it can be operated by varying the gate parameter $\delta_{0}=\epsilon_{2}-\epsilon_{1}$
or the tunneling amplitude $\tau$\citep{Martins2016} employing the
same techniques as for the regular static qubit. 

The physical origin of the energy splitting $\Delta E$ is as follows:
a magnetic field gradient reduces (increases) the energy difference
between $\ket{\dw,\up}$ ($\ket{\up,\dw}$) and the doubly occupied
states $\left\{ \ket{\up\dw,0},\ket{0,\up\dw}\right\} $. As a result,
the perturbative transition rate between $\ket{\dw,\up}$ ($\ket{\up,\dw}$)
and the doubly occupied states  is increased (reduced). This is reflected
at second order of the perturbation as a reduction of the energy splitting.
The other term appearing in Eq.~\ref{eq:HQ} is the (resonant) exchange
interaction, which corresponds to transitions between $\ket{\up,\dw}$
and $\ket{\dw,\up}$ due to virtual tunneling processes through $\mathcal{D}$.
These are represented both as a function of $\delta_{0}$ and $\epsilon_{\text{ac}}$
in Fig.~\ref{fig:0th-order} b) and~c), respectively. 

Estimating the range of validity of these expressions is particularly
difficult in the high voltage amplitude $\epsilon_{\text{ac}}/\hbar\omega\gg1$
regime. Since the sideband strength is governed by $J_{\nu}\left(\epsilon_{\text{ac}}/\hbar\omega\right)$,
there is a large number of active sidebands in this regime. Moreover,
because $\hbar\omega\sim|\Delta|\ll|\delta_{1}|,|\delta_{2}|$, there
is a certain $\tilde{\nu}$ for which at least one denominator in
Eqs.~\ref{eq:ERWA} and~\ref{eq:JRWA} will be zero or very close
to it (i.e. $\delta_{k}\pm\Delta\pm\tilde{\nu}\hbar\omega\simeq0$,
$k=1,2$). This corresponds to a photoassisted resonance with the
doubly occupied states and the perturbative expansion fails. For that
reason, we consider ac voltage amplitudes that are small enough that
we can neglect photoassisted transitions to $\mathcal{D}$. This is
the case provided that $\tau^{2}J_{\tilde{\nu}}^{2}\left(\epsilon_{\text{ac}}/\hbar\omega\right)\left(\delta_{k}\pm\Delta\pm\tilde{\nu}\hbar\omega\right)^{-1}\ll\Delta E,\mathcal{J}_{n},$
with $k=1,2$. In particular, in Appendix~\ref{sec:The-small-ac-bias}
we discuss the limit of $\epsilon_{\text{ac}}\ll\hbar\omega$, where
the expressions of Eqs.~\ref{eq:ERWA} and~\ref{eq:JRWA} have simple
forms. This is represented schematically in Fig.~\ref{fig:0th-order}~a)
where we have plotted the minimum of the denominators of Eqs.~\ref{eq:ERWA}
and~\ref{eq:JRWA} for each sideband, as well as the corresponding
Bessel function. Above a certain $\nu$ (around 40 in this case, quite
smaller than $\tilde{\nu}$ for $\delta_{0}\simeq0$) the sidebands
can be neglected and the cotunneling approximation is valid (See Appendix~\ref{sec:Beyond-the-RWA}).

Another possible route to manipulate the qubit is by employing an
ac driving in the tunneling amplitude. From the choice of gate described
below Eq.~\ref{eq:Ht}, this corresponds to replacing $J_{\nu}\left(\epsilon_{\text{ac}}/\hbar\omega\right)$
by $\tau_{\nu}$ in Eqs.~\ref{eq:interaction-tun}-\ref{eq:photoassisted-path}
of the appendix, where $\tau_{\nu}=\delta_{\nu0}\tau+\delta_{|\nu|,1}\tau_{\text{ac}}/2$.
For this simple choice of the tunneling amplitude, only resonances
with one or two photons yield a time-independent Hamiltonian like
Eq.~\ref{eq:HQ}. For $n=1$ (single-photon resonance) we have

\begin{align}
	\Delta E=2\Delta & -\tau^{2}\left[\frac{2\Delta}{\delta_{2}^{2}-\Delta^{2}}+\frac{2\Delta}{\delta_{1}^{2}-\Delta^{2}}\right]\nonumber \\
	& -\frac{\tau_{\text{ac}}^{2}}{2}\left[\frac{\Delta+\hbar\omega}{\delta_{2}^{2}-\left(\Delta+\hbar\omega\right)^{2}}+\frac{\Delta+\hbar\omega}{\delta_{1}^{2}-\left(\Delta+\hbar\omega\right)^{2}}\right]\nonumber \\
	& -\frac{\tau_{\text{ac}}^{2}}{2}\left[\frac{\Delta-\hbar\omega}{\delta_{2}^{2}-\left(\Delta-\hbar\omega\right)^{2}}+\frac{\Delta-\hbar\omega}{\delta_{1}^{2}-\left(\Delta-\hbar\omega\right)^{2}}\right],\label{eq:DeltaE-tac}
\end{align}
\begin{align}
	\mathcal{J}_{1}= & \tau\tau_{\text{ac}}\left[\frac{\delta_{1}}{\delta_{1}^{2}-\Delta^{2}}+\frac{\delta_{2}}{\delta_{2}^{2}-\Delta^{2}}\right.\nonumber \\
	& \left.\,\,\,\,\,+\frac{\delta_{1}}{\delta_{1}^{2}-\left(\Delta-\hbar\omega\right)^{2}}+\frac{\delta_{2}}{\delta_{2}-\left(\Delta-\hbar\omega\right)^{2}}\right].\label{eq:J-tac}
\end{align}
For $n=2$ (two-photons resonance) the exchange interaction is
\begin{align}
	\mathcal{J}_{2}=\frac{\tau_{\text{ac}}^{2}}{2} & \left[\frac{\delta_{1}}{\delta_{1}^{2}-\left(\Delta-\hbar\omega\right)^{2}}+\frac{\delta_{2}}{\delta_{2}^{2}-\left(\Delta-\hbar\omega\right)^{2}}\right].\label{eq:J-tac-2-photon}
\end{align}

\subsection{Operation under electric noise\label{subsec:Operation-under-charge}}

When operating the qubit, the DQD is subjected to electric noise coupling
to the Hamiltonian of Eq.~\ref{eq:HQ} through the dependence of
$\Delta E$ and $\mathcal{J}_{n}$ on $\delta_{1,2},\tau,$ and on
$\epsilon_{\text{ac}}$ or $\tau_{\text{ac}}$. This results in a
loss of coherence during qubit operation. The noisy environment can
be characterized by the spectral density which we assume has the form\citep{Burkard2009,Ali2014}
$\mathcal{S}\left(\Omega\right)=A/\left|\Omega\right|$, often called
$1/f$ noise. Following Ref.~\citep{Russ2016}, we will consider
both infrared $\Omega_{\text{IR}}$ and ultraviolet $\Omega_{\text{UV}}$
cutoffs in $\mathcal{S}\left(\Omega\right)$ for low and high frequencies,
respectively. 

In this work, we focus on the effect of charge noise entering through
the on-site energy of the dots, resulting in fluctuations of the
parameter $\delta_{0}$. Nonetheless, the same techniques described
below can be employed with other sources of electric noise. We consider
the effect of noise in the context of a Ramsey-like decay\citep{Ithier2005,Taylor2006}.
The dephasing time can then be estimated as\citep{Russ2016} 

\begin{align}
	T_{\varphi}^{-2} & =\frac{A}{2\hbar^{2}}\left(\frac{\partial\omega_{n}}{\partial\delta_{0}}\right)^{2}\log\gamma\nonumber \\
	& +\frac{A^{2}}{4\hbar^{2}}\left(\frac{\partial^{2}\omega_{n}}{\partial\delta_{0}^{2}}\right)^{2}\log^{2}\gamma+\mathcal{O}\left(A^{3}\right),\label{eq:Deph-time}
\end{align}
where $\gamma=\Omega_{\text{UV}}/\Omega_{\text{IR}}$ and we have
defined the qubit energy $\omega_{n}=\sqrt{\mathcal{J}_{n}^{2}+\left(\Delta E-n\hbar\omega\right)^{2}}$,
corresponding to half of the energy splitting between the two eigenstates
of the diagonalized Hamiltonian $\hat{H}_{\mathcal{Q}}^{\left(0\right)}$.
This expression is obtained in the context of the time-independent
RWA Hamiltonian, Eq.~\ref{eq:HQ}, but the terms neglected in the
RWA approximation do not contribute significantly~to dephasing in
the high frequency regime. In particular, in Appendix~\ref{sec:Dephasing-model}
we consider the contribution to the dephasing due to the neglected
sidebands and show that they do not yield an exponential decay of
the coherence for times larger than $\omega^{-1}$.

Since $A$ is often a small scale of the system, the first-order susceptibility
$\partial\omega_{n}/\partial\delta_{0}$ is of particular importance.
The points at which $\partial\omega_{n}/\partial\delta_{0}=0$ are
\emph{first-order sweetspots}, where the dephasing time is infinite
up to first order in $A$. Similarly, the points where $\partial\omega_{n}^{2}/\partial\delta_{0}^{2}=0$
is also zero are \emph{second-order sweetspots} and the dephasing
time is infinite up to second order in \emph{$A$}. Higher order sweetspots
are also possible\citep{AbadilloUriel2019} for configurations beyond
the scope of this work. 

For the undriven DQD-based qubit, there exists a first-order sweetspot
at $\delta_{1}=\delta_{2}$. At this point, the virtual transitions
to $\ket{\up\dw,0}$ and $\ket{0,\up\dw}$ interfere in such a way
that the first-order susceptibility is zero. This is still true in
the ac-driven qubit. In particular, for arbitrary $n$, we have at
$\delta_{1}=\delta_{2}$

\begin{align}
	\left.\frac{\partial\Delta E}{\partial\delta_{0}}\right|_{\delta_{1}=\delta_{2}}= & 0,\label{eq:first-susc--1}\\
	\left.\frac{\partial\mathcal{J}_{n}}{\partial\delta_{0}}\right|_{\delta_{1}=\delta_{2}}= & \tau^{2}\left[\left(-1\right)^{n}-1\right]\nonumber \\
	\times\Biggl[\sum_{\nu=1}^{\infty}J_{\nu}\left(\frac{\epsilon_{\text{ac}}}{\hbar\omega}\right) & J_{\nu+n}\left(\frac{\epsilon_{\text{ac}}}{\hbar\omega}\right)\frac{4\delta_{1}\left(\Delta+\nu\hbar\omega\right)}{\left[\delta_{1}^{2}-\left(\Delta+\nu\hbar\omega\right)^{2}\right]^{2}}\nonumber \\
	-\sum_{\nu=0}^{\infty}J_{\nu}\left(\frac{\epsilon_{\text{ac}}}{\hbar\omega}\right) & J_{\nu-n}\left(\frac{\epsilon_{\text{ac}}}{\hbar\omega}\right)\frac{4\delta_{1}\left(\Delta-\nu\hbar\omega\right)}{\left[\delta_{1}^{2}-\left(\Delta-\nu\hbar\omega\right)^{2}\right]^{2}}\Biggr].\label{eq:first-susc-}\\
	\left.\frac{\partial\omega_{n}}{\partial\delta_{0}}\right|_{\delta_{1}=\delta_{2}} & =\frac{\mathcal{J}_{n}}{\omega_{n}}\left.\frac{\partial\mathcal{J}_{n}}{\partial\delta_{0}}\right|_{\delta_{1}=\delta_{2}}.
\end{align}
For even $n$, $\partial\mathcal{J}_{n}/\partial\delta_{0}=\partial\omega_{n}/\partial\delta_{0}=0$
at $\delta_{1}=\delta_{2}$, as can be seen from the prefactor in
the expression for $\partial\mathcal{J}_{n}/\partial\delta_{0}$.
That is: the sweetspot in the static qubit survives to the dynamic
case. For odd $n$, on the other hand, $\mathcal{J}_{n}=0$ at $\delta_{1}=\delta_{2}$,
and therefore $\partial\omega_{n}/\partial\delta_{0}$ is also zero
at this point. Hence, while there is also a first-order sweetspot
for odd $n$ at $\delta_{1}=\delta_{2}$, the exchange interaction
is zero at it, rendering this point useless for operating the qubit.
In general, for arbitrary $n$, we have 
\begin{figure}
	\begin{centering}
		\includegraphics[width=1\columnwidth]{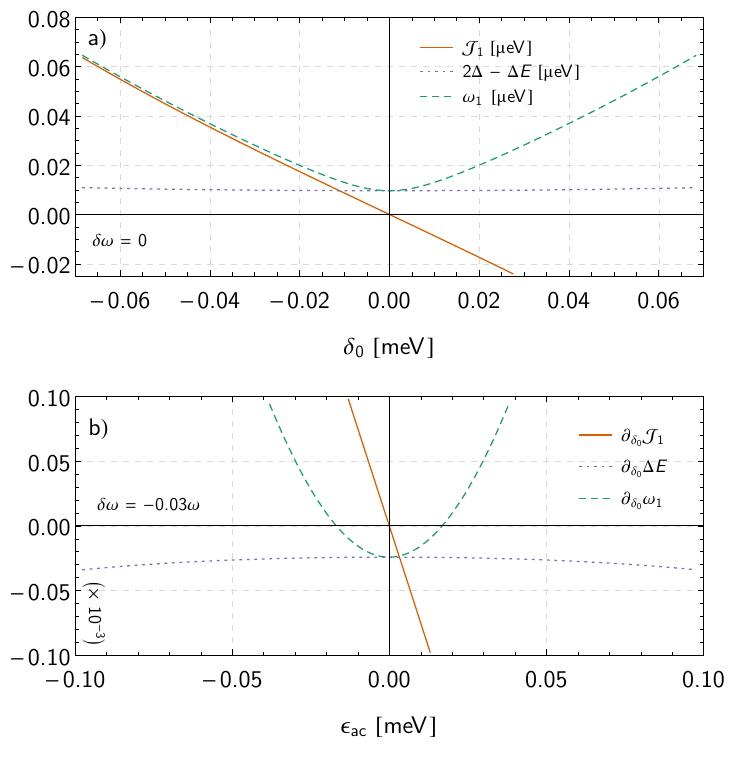}
		\par\end{centering}
	\caption{a) The exchange interaction $\mathcal{J}_{1}$ (red, continuous),
		correction to the energy splitting $2\Delta-\Delta E$ (blue, dotted),
		and qubit energy $\omega_{1}=\sqrt{\mathcal{J}_{1}^{2}+\left(\Delta E-\hbar\omega\right)^{2}}$
		(green, dashed) as a function of $\delta_{0}$ at $\epsilon_{\text{ac}}=0.12\,\text{meV}$
		with on-site driving. The exchange interaction is zero at  $\delta_{0}=0$.
		b) First order derivative of the same three quantities as a function
		of $\epsilon_{\text{ac}}$ for the same quantities and $\delta_{0}=0.07\,\text{meV}$.
		A first order sweetspot where $\partial\omega_{1}/\partial\delta_{0}=0$
		is induced by the ac voltage at $\epsilon_{\text{ac}}\simeq\pm0.07\,\text{meV}$.
		Same parameters as Fig.~\ref{fig:0th-order} with an offset of the
		resonance $\delta\omega=-0.003\omega$ in b).\label{fig:n1-resonance}}
\end{figure}

\begin{align}
	\left.\mathcal{J}_{n}\right|_{\delta_{1}=\delta_{2}}= & \tau^{2}\delta_{1}\left[\left(-1\right)^{n}+1\right]\nonumber \\
	\times & \left[\sum_{\nu=0}^{\infty}\frac{J_{\nu}\left(\frac{\epsilon_{\text{ac}}}{\hbar\omega}\right)J_{\nu+n}\left(\frac{\epsilon_{\text{ac}}}{\hbar\omega}\right)}{\delta_{1}^{2}-\left(\Delta+\nu\hbar\omega\right)^{2}}\right.\nonumber \\
	+ & \left.\sum_{\nu=1}^{\infty}\frac{J_{\nu}\left(\frac{\epsilon_{\text{ac}}}{\hbar\omega}\right)J_{\nu-n}\left(\frac{\epsilon_{\text{ac}}}{\hbar\omega}\right)}{\delta_{1}^{2}-\left(\Delta-\nu\hbar\omega\right)^{2}}\right],\label{eq:exchange-at-ss}
\end{align}
We have represented $\mathcal{J}_{1}$, the correction to the energy
splitting $2\Delta-\Delta E$ and $\omega_{1}$ in Fig.~\ref{fig:n1-resonance}~a)
as a function of $\delta_{0}$ and for $n=1$. Note that since $\mathcal{J}_{1}$
changes sign as it crosses $\delta_{0}=0$ (corresponding to $\delta_{1}=\delta_{2}$
in the symmetric configuration $U_{1}=U_{2}$), a combined manipulation
of the ac and dc gates can be employed to reverse the ground state
of the singlet-triplet qubit, enabling initialization in the triplet
subspace (in the rotating frame). Fig.~\ref{fig:n1-resonance}~b)
shows the first-order derivatives $\partial_{\delta_{0}}X$, $X=\mathcal{J}_{1},\,\Delta E,$
and~$\omega_{1}$ as a function of the ac voltage amplitude for $\delta_{0}=0.07\,\text{meV}$. 

\begin{figure}
	\begin{centering}
		\includegraphics[width=1\columnwidth]{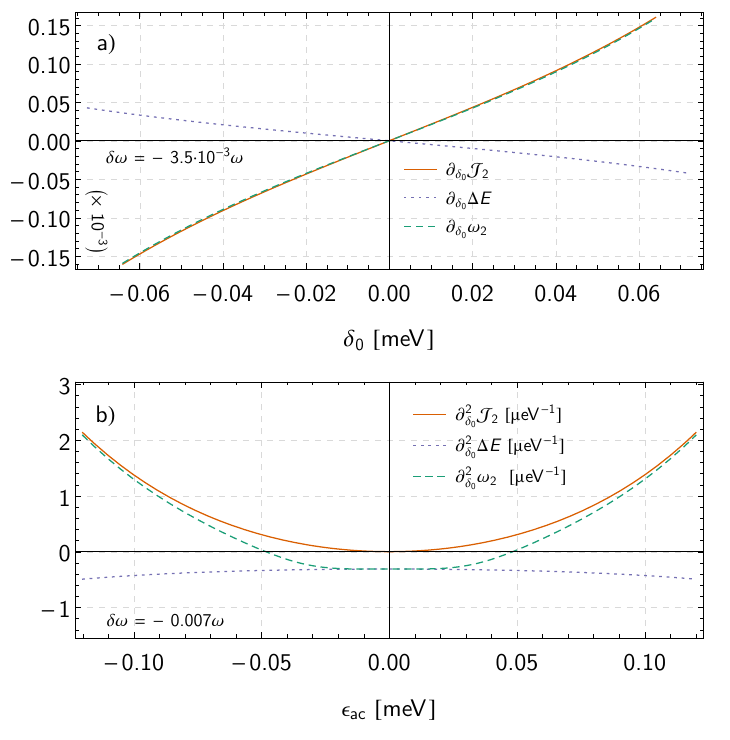}
		\par\end{centering}
	
	\caption{a) First and b) second order derivatives for the exchange interaction
		$\mathcal{J}_{2}$ (red, continuous), renormalized splitting $\Delta E$
		(blue, dotted) and the qubit splitting $\omega_{2}$ (green, dashed)
		for the two photon ($n=2$) resonance as a function of $\delta_{2}$
		and $\epsilon_{\text{ac}}$, respectively, for the case of on-site
		driving. The dc sweetspot at $\delta_{0}=0$ survives the presence
		of an ac voltage in this case, while there is a new dynamical second-order
		sweetspot where $\partial^{2}\omega_{2}/\partial\delta_{0}^{2}=0$
		at $\epsilon_{\text{ac}}\simeq\pm0.095\,\text{meV}$. Same parameters
		as in Fig.~\ref{fig:0th-order} with an offset of the resonance $\delta\omega=-0.007\omega$.
		\label{fig:1st-2nd-order}}
\end{figure}

The first-order derivatives for $n=2$, $\partial_{\delta_{0}}X$,
$X=\mathcal{J}_{2},\,\Delta E,\,\omega_{2}$, are represented in Fig.~\ref{fig:1st-2nd-order}~a),
showing the presence of a sweetspot at $\delta_{0}=0$. One particular
benefit of employing the two-photons resonance is that the system
can be kept at the first-order sweetspot while the ac gate can be
employed to increase the strength of the exchange interaction, yielding
faster operation times while still keeping the system at the sweetspot.
While this can certainly be beneficial for the qubit operation, fast
operation requires larger ac voltage amplitudes ($\propto\epsilon_{\text{ac}}^{2}$)
than for $n=1$ ($\propto\epsilon_{\text{ac}}$); moreover, the exchange
for $n$ even is $\propto\Delta^{2}$, compared to the $n$ odd, which
is $\propto\Delta$ (See Appendix~\ref{sec:The-small-gradient}). 

Once at the sweetspot, the leading source of decoherence is through
the second-order susceptibility $\partial^{2}\omega_{n}/\partial\delta_{0}^{2}$.
In the regular sweetspot $\delta_{1}=\delta_{2}$ for the static DQD
qubit, $\partial^{2}\omega_{n}/\partial\delta_{0}^{2}\neq0$, and
the second-order contribution to dephasing is dominant. In the ac-driven
DQD qubit, however, the second-order susceptibility can be zero for
finite $\epsilon_{\text{ac}}$. For instance, consider the $n=2$
resonance at $\delta_{1}=\delta_{2}$, so that $\partial\mathcal{J}_{2}/\partial\delta_{0},\partial\Delta E/\partial\delta_{0}=0$.
Then, the second order susceptibility fulfills
\begin{equation}
	\frac{\partial^{2}\omega_{2}}{\partial\delta_{0}^{2}}=\frac{1}{\omega_{2}}\left[\mathcal{J}_{2}\frac{\partial^{2}\mathcal{J}_{2}}{\partial\delta_{0}^{2}}+\left(\Delta E-2\hbar\omega\right)\frac{\partial^{2}\Delta E}{\partial\delta_{0}^{2}}\right].
\end{equation}
For a small offset of the resonance, such that $\Delta E-2\hbar\omega=\hbar\delta\omega$,
the contributions to $\partial^{2}\omega_{2}/\partial\delta_{0}^{2}$
coming from $\mathcal{J}_{2}$ and $\Delta E$ have opposite sign
and the susceptibility can be zero. This non-zero offset of the resonance
does not break the conditions for the RWA provided that $\delta\omega\ll\omega$
Then, the ac voltage amplitude can be employed to tune the susceptibility
to induce a second-order sweetspot. This renders the dephasing time
infinite under the approximation of Eq.~\ref{eq:Deph-time}. In Fig.~\ref{fig:1st-2nd-order}~b)
we show an example of this for $n=2$, where a second-order sweetspot
can be obtained at $\epsilon_{\text{ac}}\simeq0.05\,\text{meV}$.
Similarly, for $n=1$ it is possible to find an easily accessible
first-order dynamical sweetspot where $\partial\omega_{1}/\partial\delta_{0}=0$
for a small value of $\epsilon_{\text{ac}}$, as shown in Fig.~\ref{fig:n1-resonance}~b). 

\begin{figure*}
	\begin{centering}
		\includegraphics[width=2\columnwidth]{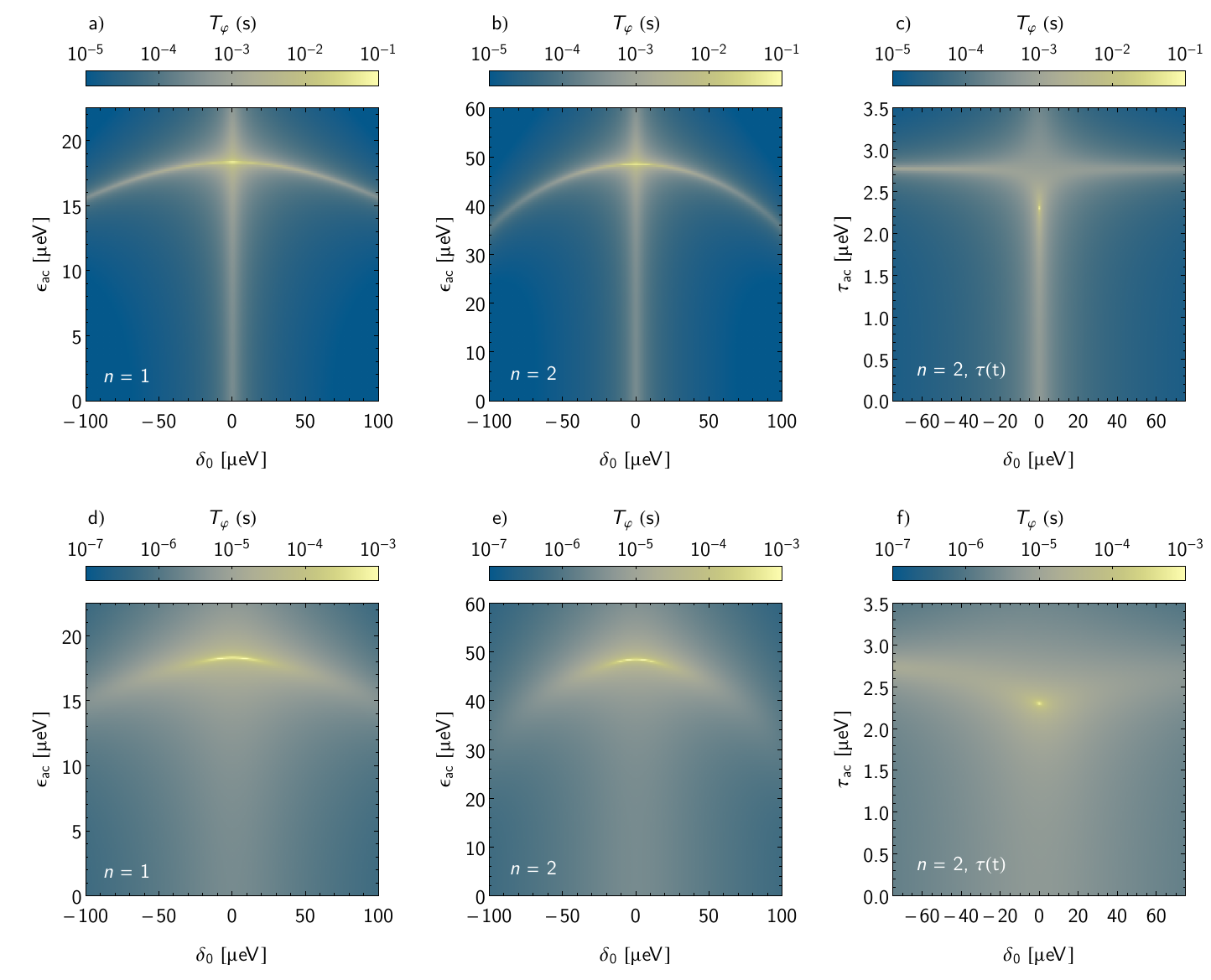}
		\par\end{centering}
	\caption{Dephasing time for $\gamma=5\cdot10^{6}$ and~$A=10^{-6}\,\text{meV}^{2}$
		(First row, Figs. a) - c)) and $A=10^{-4}\,\text{meV}^{2}$ (Second
		row, Figs. d) - f)). The first column (Figs a) and d)) correspond
		to gate driving with a one-photon resonance ($n=1$), the second column
		(Figs. b) and e)) to gate driving with a two-photon resonance ($n=2$)
		and the third column (Figs. c) and f)) to tunnel barrier driving with
		a two-photon resonance ($n=2$). Parameters: $\Delta=1.66\bar{6}\,\mu\text{eV},$
		$U_{1}=U_{2}=0.5\,\unit{meV}$, $\tau_{0}=0.01\,\text{meV}$, $V=0.1\,\unit{meV}$,
		$\gamma=5\cdot10^{6}$. The resonance offset is $\delta\omega=-0.007\omega,-0.03\omega,$ and~$-0.07\omega$ for the first, the second and the third columns, respectively. \label{fig:deph-time}}
\end{figure*}

Compared to the driving of the quantum dot gate voltage, driving the
tunneling amplitude leads to an exchange interaction that does not
vanish at $\delta_{1}=\delta_{2}$ for $n=1$ (and also does not change
sign depending on $\delta_{0}$). For $n=2$, $\delta_{1}=\delta_{2}$
is still a first-order sweetspot, while for $n=1$ this sweetspot
is displaced to $\delta_{0}\neq0$. Following the same technique
described in the previous section, the system can be tuned to induce
a second-order sweetspot, using in this case the ac tunneling amplitude
$\tau_{\text{ac}}$. 

In Fig.~\ref{fig:deph-time} we represent the dephasing time for
$\gamma=5\cdot10^{6}$ and~$A=10^{-6}\,\text{meV}^{2}$ (Figs.~\ref{fig:deph-time}~a)
- c)) and $A=10^{-4}\,\text{meV}^{2}$ (Figs.~\ref{fig:deph-time}~d)
- f)), corresponding to the lower and upper bounds of the realistic
range\citep{Petersson2010,Russ2015} $\sqrt{A}\sim1-10\,\mu\text{eV}$,
respectively. From left to right, these are: the single-photon and
two-photon resonances with the ac driving in the dot gates and the
two-photon resonance with the ac driving in the tunneling gates. We
have chosen an offset of the resonance condition that enables us to
induce a first-order sweetspot for $n=1$ and a second-order sweetspot
for the two configurations with $n=2$.

For the case with the ac driving in the dot gates and $n=1$, the
exchange interaction is zero at $\delta_{0}=0$, as discussed above.
Outside of $\delta_{0}=0$, in the arc around $\epsilon_{\text{ac}}\simeq15-18\,\mu\text{eV}$
(corresponding to the dynamical first-order sweetspot mentioned in
Sec.~\ref{subsec:Operation-under-charge}), the dephasing time is
still quite large, $T_{\varphi}\sim10^{-4}\,\text{s}$ for $A=10^{-6}\,\text{meV}^{2}$
(Fig.~\ref{fig:deph-time}~a)), compared with the typical Rabi period
in this configuration ($\mathcal{T}_{\Omega}\sim10^{-7}\,\text{s})$.
For a higher noise intensity (Fig.~\ref{fig:deph-time}~d)), the
region of long dephasing time corresponding to this dynamical sweetspot
is less pronounced as the second-order contribution is more important.
In the region close to the first-order dynamical sweetspot it is still
possible to achieve a dephasing time of the order of $10^{-3}\,\text{s}$,
but this decreases quickly as $\delta_{0}$ is varied from zero. As
a result, a compromise has to be reached between faster operation
(moving away from $\delta_{0}=0$) and increased dephasing time (moving
closer to $\delta_{0}=0$).

For $n=2$, the intersection of $\delta_{0}=0$ with $\epsilon_{\text{ac}}\simeq48\,\mu\text{eV}$
defines the second-order sweetspot discussed above, where the dephasing
time is infinite under the dephasing model of Eq.~\ref{eq:Deph-time}
(i.e., up to terms $\propto A^{3}$). For $A=10^{-6}\,\text{meV}^{2}$
(Fig.~\ref{fig:deph-time}~b)) the dephasing time is quite large
($T_{\varphi}\sim1-10\,\text{ms}$) in the area around the sweetspot.
The second-order sweetspot is still visible in Fig.~\ref{fig:deph-time}~e),
corresponding to higher noise intensity. However, while in Fig.~\ref{fig:deph-time}~b)
$\delta_{0}=0$ corresponds to a region of longer dephasing time,
in Fig.~\ref{fig:deph-time}~e) the normal sweetspot has decreased
in importane, while the dynamical sweetspot still yields large dephasing
times even relatively far from $\delta_{0}=0$, showing the robustness
of the dynamical sweetspot

For the case of driving in the tunneling gate and $n=2$ (Fig.~\ref{fig:deph-time}~c)
and~f)), as mentioned above, the first-order sweetspot is still at
$\delta_{0}=0$, which can be seen clearly from the long dephasing
time ($T_{\varphi}\sim10^{-5}-10^{-4}\,\text{s}$) in Fig.~\ref{fig:deph-time}~c)
and less so in Fig.~\ref{fig:deph-time}~f) due to the higher noise
intensity giving increased importance to the second-order term. Fig.~\ref{fig:deph-time}~c)
shows clearly a second-order sweetspot appearing at $\tau_{\text{ac}}\simeq5\,\mu\text{eV}$,
and $\delta_{0}=0$. Again, the results for stronger noise in Fig.~\ref{fig:deph-time}~f)
show the second-order sweetspot clearly. Interestingly, while the
region of longer dephasing time at $\delta_{0}=0$ is much less marked,
the region of longer dephasing time around $\tau_{\text{ac}}\simeq5\,\mu\text{eV}$
is still appreciable.

\section{Discussion\label{sec:Discussion}}

In this work we have studied the possibility of employing ac gates
to engineer the properties of quantum-dot based qubits under charge
noise. When quantum dot chains are driven with high ac voltage amplitudes,
many sidebands become available as virtual transitions, producing
new and interesting phenomena. The single photon resonance results
in an exchange interaction that vanishes in the sweetspot defined
in the static system and changes sign across this point. This leads
to the possibility of manipulating its sign with the combined action
of the dc and ac gates, and thus the ground state of the singlet-triplet
qubit. Moreover, we have shown how the presence of a time-dependent
voltage can produce a dynamical sweetspot, mitigating the effect of
electric noise. For the two-photon resonance, we have shown that the
ac voltage can be employed to induce a second order sweetspot, that
is, a configuration in which both the first-order and second-order
susceptibilities to electric noise are zero. In particular, we have
considered the second order susceptibility of the qubit energy in
a free decay model, showing that the dephasing time becomes infinite
up to third order in the noise strength. We have also studied an alternative
setup in which the ac driving occurs in the tunnel barrier, which
provides an alternative method of controlling the exchange interaction
in which it is also possible to find second-order sweetspots. These
setups showcase new routes to attain high-order sweetspots through
time-dependent gates, adding to previous proposals involving large
tunneling amplitudes\citep{AbadilloUriel2019}. Overall, our work shows
the possibilities of employing Floquet engineering to enhance the
robustness of qubits in electric environments which could be generalized
to other configurations, for instance for the triple quantum dot-based
resonant exchange qubit. This method is within the reach of current
experimental techniques. 
\begin{acknowledgments}
	We acknowledge Sigmund Kohler for enlightening discussions and a critical
	reading of the manuscript and support from CSIC Research Platform
	PTI-001. G.P. acknowledges her Mercator Fellow position at CRC 1277,
	University of Regensburg. This work was supported by the Spanish Ministry
	of Economy and Competitiveness (MICINN) via Grants No. PID2020-117787GB-100
	and No. MAT-2017-86717-P. 
\end{acknowledgments}

\bibliography{tqd}

\appendix

\section{Effective cotunneling Hamiltonian\label{sec:Effective-cotunneling-Hamiltonia}}

In this appendix, we obtain the effective cotunneling Hamiltonian
of Eq.~\ref{eq:HQ} starting from the full Hamiltonian for the DQD.
We will consider first the case of the ac modulation in the quantum
dot gate. In this case, the DQD Hamiltonian, $\hat{H}\left(t\right)$
(Eq.~\ref{eq:Ht}) can be written in the interaction picture with
respect to $\hat{H}_{\mathrm{ac}}\left(t\right)$ by a unitary transformation
$\hat{H}_{I}\left(t\right)=\hat{U}_{I}\left(t\right)[\hat{H}\left(t\right)-i\hbar\partial_{t}]\hat{U}_{I}^{\dagger}\left(t\right)$
where $\hat{U}_{I}\left(t\right)=\exp[i\hbar^{-1}\int\hat{H}_{\mathrm{ac}}\left(s\right)ds]$.
The transformed Hamiltonian reads $\hat{H}_{I}\left(t\right)=\hat{H}_{0}+\hat{H}_{U}+\hat{H}{}_{T,I}\left(t\right)$,
where

\begin{align}
	\hat{H}_{T,I}\left(t\right) & =\sum_{\alpha,\beta;\sigma}\sum_{\nu=-\infty}^{\infty}\tau_{\mathrm{\nu}}\left(t\right)\left(\hat{c}_{\alpha,\sigma}^{\dagger}\hat{c}_{\beta,\sigma}+\mathrm{H.c.}\right),\label{eq:interaction-tun}\\
	\tau_{\mathrm{\nu}}\left(t\right) & =\tau J_{\nu}\left(\epsilon_{\text{ac}}/\hbar\omega\right)\exp[i\nu\left(\omega t+\phi\right)].\label{eq:photoassisted-path}
\end{align}
The transformed tunneling term, $\hat{H}_{T,I}\left(t\right)$, includes
all possible sideband transitions $\tau_{\nu}\left(t\right)$ involving
$\nu$ photons, either absorbed $\left(\nu>0\right)$ or emitted $\left(\nu<0\right)$.

If the states with double occupancy are sufficiently separated in
energy from the states in $\mathcal{Q}$, we may perform a time-dependent
Schrieffer-Wolff transformation\citep{Goldin2000,Pico-Cortes2019}
(SWT) with respect to $H_{T,I}\left(t\right)$ in order to obtain
an effective Hamiltonian for the low energy subspace $\mathcal{Q}$,
with virtual tunneling as the leading order of perturbation, given
by $\hat{H}_{\text{SW}}\left(t\right)=e^{\hat{\Upsilon}(t)}\hat{H}_{I}\left(t\right)e^{-\hat{\Upsilon}(t)}-i\hbar\partial_{t}\hat{\Upsilon}\left(t\right)$,
where $\hat{\Upsilon}\left(t\right)=\sum_{\sigma}s_{\beta\alpha}\left(t\right)\hat{c}_{\alpha,\sigma}^{\dagger}\hat{c}_{\beta,\sigma}$
is an anti-Hermitian operator and the $s_{\alpha\beta}\left(t\right)$
satisfy 
\begin{align}
	i\hbar\partial_{t}s_{\beta\alpha}\left(t\right)= & \sum_{\nu=-\infty}^{\infty}\tau_{\nu}\left(t\right)-s_{\beta\alpha}\left(t\right)\left(\epsilon_{\alpha}-\epsilon_{\beta}\right).\label{eq:seq}
\end{align}
Solving these equations together with the condition that $s_{\beta\alpha}\left(t\right)$
is time-independent in the absence of external potentials we obtain
the effective Hamiltonian. Denoting the zeroth order splitting between
$\ket{\up,\dw}$ and $\ket{\dw,\up}$ as $\Delta=E_{z,1}-E_{z,2}$,
the effective Hamiltonian in the $\mathcal{Q}=\left\{ \ket{\up,\dw},\ket{\dw,\up}\right\} $
subspace is 

\begin{equation}
	\hat{H}_{\text{SW}}\left(t\right)=\begin{pmatrix}E^{+}\left(t\right) & \mathcal{J}^{*}\left(t\right)\\
		\mathcal{J}\left(t\right) & E^{-}\left(t\right)
	\end{pmatrix},
\end{equation}
where the matrix elements are

\begin{align}
	E^{\pm}\left(t\right)=\pm\Delta\label{eq:E-of-t}\\
	-2\sum_{\nu,\mu} & \left[\frac{\Re\left\{ \tau_{\nu}^{*}\left(t\right)\tau_{\mu}\left(t\right)\right\} }{\delta_{2}\mp\Delta-\nu\hbar\omega}+\frac{\Re\left\{ \tau_{\nu}^{*}\left(t\right)\tau_{\mu}\left(t\right)\right\} }{\delta_{1}\mp\Delta+\nu\hbar\omega}\right],\nonumber \\
	\mathcal{J}\left(t\right)=\sum_{\mu,\nu} & \left[\frac{\tau_{\mu}^{*}\left(t\right)\tau_{\nu}\left(t\right)}{\delta_{1}-\Delta+\nu\hbar\omega}+\frac{\tau_{\mu}^{*}\left(t\right)\tau_{\nu}\left(t\right)}{\delta_{2}+\Delta-\nu\hbar\omega}\right.\nonumber \\
	+ & \left.\frac{\tau_{\nu}^{*}\left(t\right)\tau_{\mu}\left(t\right)}{\delta_{1}+\Delta+\nu\hbar\omega}+\frac{\tau_{\nu}^{*}\left(t\right)\tau_{\mu}\left(t\right)}{\delta_{2}-\Delta-\nu\hbar\omega}\right].\label{eq:J-of-t}
\end{align}
The exchange interaction is a second-order process. As a result, when
an ac gate is introduced into the system, the time-dependent rate
$\mathcal{J}\left(t\right)$ involves two sideband transitions with
$\mu$ and $\nu$ photons, respectively. $E^{\pm}\left(t\right)$
incorporates the gradient splitting between $\ket{\up,\dw}$ and $\ket{\dw,\up}$
and a time-dependent gradient (similar to the effective ac magnetic
field in EDSR). As expected, $E^{\pm}\left(t\right)\to0$ as $\Delta\to0$,
as in that case the SU(2) symmetry between the two qubit states is
not broken. The case with the ac modulation in the tunneling can
be obtained from this expression by taking only the sidebands with
$\mu,\nu=0,\pm1$. Moreover, we note that, if both modulations of
the tunneling and gate energies were present, the transformed Hamiltonian
of Eq.~\ref{eq:interaction-tun} would include two extra contributions
of the form
\begin{align}
	\hat{H}_{T,I}^{\prime}\left(t\right) & =\sum_{\alpha,\beta;\sigma}\sum_{\nu}\left[\tau_{\mathrm{\nu}}^{+}\left(t\right)+\tau_{\mathrm{\nu}}^{-}\left(t\right)\right]\left(\hat{c}_{\alpha,\sigma}^{\dagger}\hat{c}_{\beta,\sigma}+\mathrm{H.c.}\right),\label{eq:interaction-tun-1}\\
	\tau_{\mathrm{\nu}}^{\pm}\left(t\right) & =(\tau_{\text{ac}}/2)J_{\nu}\left(\epsilon_{\text{ac}}/\hbar\omega\right)\exp[i\left(\nu\pm1\right)\left(\omega t+\phi\right)].\label{eq:photoassisted-path-1}
\end{align}
As a result, compared to the previous case, each sideband would be
further split into three, one from Eq.~\ref{eq:interaction-tun}
and the two from Eq.~\ref{eq:interaction-tun-1}.

Finally, in order to obtain the RWA Hamiltonian, a unitary transformation
to the rotating frame is performed, $\hat{H}_{n}\left(t\right)=\hat{U}_{n}\left(t\right)[\hat{H}_{\text{SW}}\left(t\right)-i\hbar\partial_{t}]\hat{U}_{n}^{\dagger}\left(t\right)$,
where $\hat{U}_{n}\left(t\right)=\exp[in\omega t(\hat{S}_{z,1}-\hat{S}_{z,2})/\hbar]$
and $n\mathcal{2}\mathbb{Z}$. The two states are resonant provided
that
\begin{align}
	\frac{1}{T}\int_{0}^{T}dt\left(\bra{\up,\dw}\hat{H}_{n}\left(t\right)\ket{\up,\dw}\right.\nonumber \\
	\left.-\bra{\dw,\up}\hat{H}_{n}\left(t\right)\ket{\dw,\up}\right) & =n\hbar\omega,\,\,\,n\mathcal{2}\mathbb{Z},
\end{align}
where $T=2\pi/\omega$. The integer $n$ denotes the order of the
resonance and represents the number of photons involved in the resonant
transition. The condition can be written as $\Delta E=E^{+}-E^{-}=n\hbar\omega$,
where the mean energies $E^{\pm}$ in the RWA are given by Eq.~\ref{eq:ERWA}.
Then, as mentioned in the main text, the oscillatory components can
be neglected.

\section{Corrections to the approximation\label{sec:Beyond-the-RWA}}

As mentioned in the main text, the SWT-based expansion up to terms
$\propto\tau^{2}$ is valid provided that the tunneling-widened levels
are not resonant, i.e. provided that $\tau/(\delta_{1}\pm\Delta\pm\tilde{\nu}\hbar\omega)\ll1$.
In particular, Eqs.~\ref{eq:E-of-t}-\ref{eq:J-of-t} are valid up
to factors $\propto\tau^{4}$. The next correction to $\hat{H}_{\text{SW}}(t)$
is $(1/2)[\hat{\Upsilon}(t),[\hat{\Upsilon}(t),\hat{H}_{T}(t)]]$
(which is $\propto\tau^{3}$). This term is zero when projected to
the qubit subspace $\mathcal{Q}$, so that the next order correction
to $\hat{H}_{\text{SW}}(t)$ is $\propto\tau^{4}$.

We may consider also what happen for larger ac voltage amplitudes
(i.e. breaking the conditions detailed at the end of Sec.~\ref{subsec:Resonant-operation}).
Eventually, the denominators of Eq.~\ref{eq:ERWA} and~\ref{eq:JRWA}
become negative for large $\nu\,(>\tilde{\nu})$. For large ac voltage
amplitudes $\epsilon_{\text{ac}}$, several sidebands with positive
or negative denominators may interfere. This could produce a new set
of dynamical sweetspots at large values of the ac voltage, in a similar
way to the case of Ref.~\citep{Pico-Cortes2019}. However, for the
parameters considered here this results in a set of sidebands which
are almost resonant, breaking the conditions for the cotunneling approximation.
This results in a strong hybridization with the states in $\mathcal{D}$,
which increases the effect of the electric fluctuations. 

With respect to the RWA, the time-independent exchange rates, as obtained
in Appendix~\ref{sec:Effective-cotunneling-Hamiltonia}, are valid
provided that $|\mathcal{J}_{n}|/\hbar\omega\ll1$. This corresponds
to a lowest-order approximation in a perturbative series in $1/\omega$,
and its limitations have been studied in detail elsewhere\citep{Song2016}.
Here, we follow the high frequency expansion described in Ref.~\citep{Eckardt2015}.
After applying $\hat{U}_{n}(t)$ and switching to the rotating frame\footnote{The perturbative theory of Ref.~\citep{Eckardt2015} provides an
	alternative understanding for the unitary transformation $\hat{U}_{n}(t)$
	described in Appendix~\ref{sec:Effective-cotunneling-Hamiltonia}.
	The necessary condition to develop the perturbative high-frequency
	expansion described therein is that $\hbar\omega$ is the largest
	energy scale of the system, which is accomplished by removing the
	splitting through $\hat{U}_{n}(t)$.}, the Hamiltonian can be written as a Fourier series $\hat{H}_{n}(t)=\sum_{m}e^{im\omega t}\hat{H}_{n,m}$,
where the Hamiltonian of Eq.~\ref{eq:HQ} is $\hat{H}_{\mathcal{Q}}^{\left(0\right)}=\hat{H}_{n,0}$.
We then have $\hat{P}_{\mathcal{Q}}\hat{H}_{n,m}\hat{P}_{\mathcal{Q}}^{-1}=\Delta E_{m}^{\left(0\right)}\hat{\sigma}_{z}+\Re\{\mathcal{J}_{n,m}^{\left(0\right)}\}\hat{\sigma}_{x}+\Im\{\mathcal{J}_{n,m}^{\left(0\right)}\}\hat{\sigma}_{y}$,
where $\hat{P}_{\mathcal{Q}}$ is the projector to the qubit subspace
$\mathcal{Q}$, $\Delta E_{m}^{\left(0\right)}$ and $\mathcal{J}_{n,m}$
are the $m$th harmonics of the splitting and the exchange, respectively,
from Eq.~\ref{eq:E-of-t} and~\ref{eq:J-of-t} after applying $\hat{U}_{n}\left(t\right)$.
Then the next order correction to the RWA approximation, Eq.~\ref{eq:HQ},
is given by
\begin{align}
	\hat{H}_{\mathcal{Q}}^{\left(1\right)} & =\sum_{m\neq0}\frac{\hat{H}_{n,m}\hat{H}_{n,-m}}{m\hbar\omega}\nonumber \\
	& =\left[\Delta E_{n}^{\left(1\right)}\hat{\sigma}_{z}+\Re\left\{ \mathcal{J}_{n}^{\left(1\right)}\right\} \hat{\sigma}_{x}+\Im\left\{ \mathcal{J}_{n}^{\left(1\right)}\right\} \hat{\sigma}_{y}\right]\label{eq:high-freq-expansion}
\end{align}
where
\begin{align}
	\Delta E_{n}^{\left(1\right)} & =\sum_{m=1}^{\infty}\frac{1}{2m\hbar\omega}\left(\left|\mathcal{J}_{n,-m}^{\left(0\right)}\right|^{2}-\left|\mathcal{J}_{n,m}^{\left(0\right)}\right|^{2}\right),\\
	\mathcal{J}_{n}^{\left(1\right)} & =\sum_{m=1}^{\infty}\frac{1}{m\hbar\omega}\left(\mathcal{J}_{n,m}^{\left(0\right)}\Delta E_{m}^{\left(0\right)}-\mathcal{J}_{n,-m}^{\left(0\right)}\Delta E_{m}^{\left(0\right)}\right).
\end{align}
The leading terms of this correction are $\propto\tau^{4}/\omega$.
Since, $\delta_{1},\delta_{2}\gg\hbar\omega$, this is often of greater
interest than the corrections due to the SWT. The strength of these
contributions can be estimated based on these expressions to be $\simeq|\mathcal{J}_{n}|/\hbar\omega\simeq10^{-3}$
lower than the terms in Eq.~\ref{eq:HQ}. In any case, as seen in
Eq.~\ref{eq:high-freq-expansion}, the new terms in the expansion
do not alter the way that the protocol operates. 

\section{The limit of small ac voltage amplitude\label{sec:The-small-ac-bias}}

In this appendix, we briefly describe the $\epsilon_{\text{ac}}/\hbar\omega\ll1$
limit. We proceed by linearizing the expressions for $\mathcal{J}_{1}$
and $\Delta E$ (Eqs.~\ref{eq:JRWA} and~\ref{eq:ERWA}, respectively).
Since $J_{\nu}\left(z\right)\simeq\left(z/2\right)^{\nu}/\Gamma\left(\nu+1\right)$
for $z\ll1$, where $\Gamma\left(z\right)$ is Euler's gamma function,
we see that the next terms in the splitting are $\mathcal{O}(\epsilon_{\text{ac}}^{2})$.
Hence, the result is equivalent to the dc case
\begin{align}
	\Delta E & \simeq2\Delta\left[1-\frac{2\tau^{2}}{\delta_{2}^{2}-\Delta^{2}}+\frac{2\tau^{2}}{\delta_{1}^{2}-\Delta^{2}}\right],
\end{align}
Regarding the exchange interaction, the linearized expression reads

\begin{align}
	& \mathcal{J}_{1}\simeq\frac{\tau^{2}\epsilon_{\text{ac}}}{2\hbar\omega}\nonumber \\
	\times & \left(\frac{1}{\delta_{1}-\Delta}+\frac{1}{\delta_{2}+\Delta}+\frac{1}{\delta_{2}-\Delta+\hbar\omega}+\frac{1}{\delta_{1}+\Delta-\hbar\omega}\right.\nonumber \\
	- & \left.\frac{1}{\delta_{2}-\Delta}-\frac{1}{\delta_{1}+\Delta}-\frac{1}{\delta_{1}-\Delta+\hbar\omega}-\frac{1}{\delta_{2}+\Delta-\hbar\omega}\right).\label{eq:low-eac-limit}
\end{align}
Taking the resonance condition to correspond exactly to $\hbar\omega=2\Delta$
yields the simpler expression

\begin{align}
	\mathcal{J}_{1}\simeq & \epsilon_{\text{ac}}\tau^{2}\left(\frac{1}{\delta_{1}^{2}-\Delta^{2}}-\frac{1}{\delta_{2}^{2}-\Delta^{2}}\right).\label{eq:low-eac-limit-1-1}
\end{align}
In this limit, the RWA corresponds to ignoring (1) the dc component
of $\mathcal{J}\left(t\right)$ and (2) the counter-rotating term.
The remainder of the exchange interaction is 

\begin{align}
	e^{-i\omega t}\mathcal{J}\left(t\right)-\mathcal{J}_{1}\simeq2e^{-i\omega t}\tau^{2} & \left(\frac{\delta_{1}}{\delta_{1}^{2}-\Delta^{2}}+\frac{\delta_{1}}{\delta_{2}^{2}-\Delta^{2}}\right)\nonumber \\
	+\frac{\epsilon_{\text{ac}}\tau^{2}}{\hbar\omega}e^{-i2\omega t}e^{-i\phi} & \left[\frac{\Delta}{\delta_{1}^{2}-\Delta^{2}}-\frac{\Delta}{\delta_{2}^{2}-\Delta^{2}}\right.\nonumber \\
	-\frac{\Delta-\hbar\omega}{\delta_{1}^{2}-\left(\Delta+\hbar\omega\right)^{2}} & \left.+\frac{\Delta-\hbar\omega}{\delta_{2}^{2}-\left(\Delta+\hbar\omega\right)^{2}}\right].\label{eq:small-ac-exch-cor}
\end{align}
Applying Eq.~\ref{eq:high-freq-expansion} in order to correct for
these neglected terms yields a renormalization of the splitting, $\hat{H}_{\mathcal{Q}}^{\left(1\right)}\simeq\Delta E_{1}^{\left(1\right)}\ket{\up,\dw}\bra{\up,\dw}$,
where
\begin{align}
	\Delta E_{1}^{\left(1\right)}= & -\frac{4\tau^{4}}{\hbar\omega}\left(\frac{\delta_{1}}{\delta_{1}^{2}-\Delta^{2}}+\frac{\delta_{1}}{\delta_{2}^{2}-\Delta^{2}}\right)^{2}\nonumber \\
	& -\frac{\epsilon_{\text{ac}}^{2}\tau^{4}}{\hbar^{3}\omega^{3}}\left[\frac{\Delta}{\delta_{1}^{2}-\Delta^{2}}-\frac{\Delta}{\delta_{2}^{2}-\Delta^{2}}\right.\nonumber \\
	& -\frac{\Delta-\hbar\omega}{\delta_{1}^{2}-\left(\Delta+\hbar\omega\right)^{2}}\left.+\frac{\Delta-\hbar\omega}{\delta_{2}^{2}-\left(\Delta+\hbar\omega\right)^{2}}\right]^{2}.\label{eq:small-ac-split-cor}
\end{align}
The counter-rotating terms and are $\propto\epsilon_{\text{ac}}^{2}$
and as such can be neglected in this limit. 

\section{The limit of small gradient \label{sec:The-small-gradient}}

In this section we describe the small gradient limit $|\Delta|,|\nu|\hbar\omega\ll\delta_{1},\delta_{2}$,
for all $\nu$ which are not negligible as discussed at the end of
Sec.~\ref{subsec:Resonant-operation}. This is often a good approximation,
since the interaction strength is the largest energy scale of the
system. In that case, we can linearize the splitting (Eq.~\ref{eq:ERWA})
to give 

\begin{align*}
	\Delta E & \simeq\Delta+2\tau^{2}\sum_{\nu=-\infty}^{\infty}J_{\nu}^{2}\left(\frac{\epsilon_{\text{ac}}}{\hbar\omega}\right)\left[\frac{1}{\delta_{2}}+\frac{1}{\delta_{1}}-\frac{1}{\delta_{2}}-\frac{1}{\delta_{1}}\right.\\
	& \left.-\frac{\Delta-\nu\hbar\omega}{\delta_{2}^{2}}-\frac{\Delta+\nu\hbar\omega}{\delta_{1}^{2}}-\frac{\Delta+\nu\hbar\omega}{\delta_{2}^{2}}-\frac{\Delta-\nu\hbar\omega}{\delta_{1}^{2}}\right].
\end{align*}
The terms independent of $\Delta$ are clearly zero, while the other
terms are independent of the frequency, so that the Bessel function
is the only factor that depends on the sideband index. Hence, we can
sum over $\nu$ to yield 
\begin{align}
	\Delta E & \simeq2\Delta\left[1-2\tau^{2}\left(\frac{1}{\delta_{2}^{2}}+\frac{1}{\delta_{1}^{2}}\right)\right],
\end{align}
where we have employed that $\sum_{\nu=1}^{\infty}J_{\nu}^{2}\left(z\right)=\left(1/2\right)\left[1-J_{0}^{2}\left(z\right)\right]$.
This expression is independent of the ac voltage amplitude and frequency.
Regarding the exchange interaction, we have

\begin{align}
	& \mathcal{J}_{n}\simeq\tau^{2}\left[\left(-1\right)^{n}-1\right]\left(\frac{1}{\delta_{1}^{2}}-\frac{1}{\delta_{2}^{2}}\right)\nonumber \\
	& \times\left[\sum_{\sigma=\pm1}\sum_{\nu=\delta_{\sigma1}}^{\infty}J_{\nu}\left(\frac{\epsilon_{\text{ac}}}{\hbar\omega}\right)J_{\nu+\sigma n}\left(\frac{\epsilon_{\text{ac}}}{\hbar\omega}\right)\left(\Delta+\sigma\nu\hbar\omega\right)\right].\label{eq:exchange-low-grad-1}
\end{align}
As mentioned in the main text, for low gradients the odd order resonances
are $\propto\Delta$ while the resonances with $n$ even are $\propto\Delta^{2}$. 

\section{Dephasing model\label{sec:Dephasing-model}}

We consider pure dephasing in the context of a Ramsey experiment\citep{Makhlin2004,Ithier2005}.
We focus on the function
\begin{equation}
	f\left(t\right)=\left\langle \exp\left(i\delta\hat{\varphi}\left(t\right)\right)\right\rangle =\exp\left(\frac{-\Delta\varphi\left(t\right)}{2}\right),\label{eq:decoherence-f}
\end{equation}
which describes the decay of coherence under the effect of longitudinal
noise. Here, $\delta\hat{\varphi}\left(t\right)=\hbar^{-1}\sum_{ij}\int_{0}^{t}ds\lambda_{i}\left(s\right)\hat{\xi}_{j}\left(s\right)$
is the accumulated phase due to a random variation of the bath operator
$\hat{\xi}_{j}\left(t\right)$ coupled to the system through $\lambda_{i}\left(t\right)$,
$\left\langle \cdots\right\rangle =\text{Tr}_{B}\left\{ \cdots\hat{\rho}_{B}\right\} $,
with $\hat{\rho}_{B}$ the equilibrium density matrix of the bath
and $\text{Var}[x]$ is the variance. The second equality in Eq.~\ref{eq:decoherence-f}
is obtained after expanding $f\left(t\right)$ in terms of its cumulants\citep{Russ2015}
and assuming Gaussian noise with zero mean. We can then write $\Delta\varphi\left(t\right)$
as a series in powers of $A$ as
\begin{equation}
	\Delta\varphi\left(t\right)=\Delta\varphi^{\left(1\right)}\left(t\right)+\Delta\varphi^{\left(2\right)}\left(t\right)+\ldots
\end{equation}
For the first-order coupling $\Delta\varphi^{\left(1\right)}\left(t\right)=\text{Var}\left[\delta\hat{\varphi}\left(t\right)\right]=\left\langle \delta\hat{\varphi}^{2}\left(t\right)\right\rangle $.
We consider first the dephasing due to the RWA Hamiltonian of Eq.~\ref{eq:HQ}
\begin{align}
	\Delta\varphi^{\left(1\right)}\left(t\right)\simeq & \frac{t^{2}}{\hbar^{2}}\left(\frac{\partial\omega_{n}}{\partial\delta_{0}}\right)^{2}\int_{-\infty}^{\infty}d\Omega\mathcal{S}\left(\Omega\right)\text{sinc}^{2}\left(\frac{\Omega t}{2}\right).\nonumber \\
	\simeq & \frac{At^{2}}{2\hbar^{2}}\left(\frac{\partial\omega_{n}}{\partial\delta_{0}}\right)^{2}\log\left(\frac{\Omega_{\text{UV}}}{\Omega_{\text{IR}}}\right).\label{eq:accumul-phase}
\end{align}
This is exactly as considered in Eq.~\ref{eq:Deph-time} and equivalent
to the free decay in the case without driving\citep{Russ2015}. Next,
we consider also all the other sidebands in Eqs.~\ref{eq:E-of-t}
and~\ref{eq:J-of-t} (i.e. all $\nu$ and $\mu$). These yield\begin{widetext}
	
	\begin{align}
		\Delta\varphi^{\left(1\right)}\left(t\right) & \simeq\sum_{m,m'\neq0}\frac{\partial F_{n,m}}{\partial\delta_{0}}\frac{\partial F_{n,m'}}{\partial\delta_{0}}\int_{0}^{t}dse^{im\omega s}\int_{0}^{t}ds'e^{im'\omega s'}\left\langle \hat{\xi}_{i}\left(s\right)\hat{\xi}_{i}\left(s'\right)\right\rangle .\label{eq:accumul-phase-ac-2}
	\end{align}
	If we consider that the bath dynamics are quasistatic with respect
	to the ac voltage dynamics, the bath operators in Eq.~\ref{eq:accumul-phase-ac-2}
	can be assumed to be constant in time and extracted from the integral.
	The resulting contribution oscillates in time without decaying. This
	essentially divides the contribution from the oscillating terms into
	a low frequency part which does not contribute to dephasing and a
	high frequency part which describes the time scales related to the
	micromotion (i.e. within the ac voltage period). In particular, the
	contribution $m'=-m$ acquires a form that is reminiscent of Eq.~\ref{eq:accumul-phase}
	
	\begin{equation}
		\Delta\varphi^{\left(1\right)}\left(t\right)\simeq\frac{t^{2}}{\hbar^{2}}\sum_{m\neq0}\frac{\partial F_{n,m}}{\partial\delta_{0}}\frac{\partial F_{n,\bar{m}}}{\partial\delta_{0}}\int_{-\infty}^{\infty}d\Omega C\left(\Omega-m\omega\right)\text{sinc}^{2}\left(\frac{\Omega t}{2}\right),\label{eq:accumul-phase-ac-1}
	\end{equation}
	where $C\left(\Omega\right)=\mathcal{S}\left(\Omega\right)+\mathcal{A}\left(\Omega\right)$
	is the correlation function of the bath, with symmetric $\mathcal{S}\left(\Omega\right)$
	and antisymmetric $\mathcal{A}\left(\Omega\right)$ components, and
	$F_{n,m}=\omega_{n}^{-1}[\Delta E_{m}^{\left(0\right)}\left(\Delta E-n\hbar\omega\right)+\mathcal{J}_{n,m}^{\left(0\right)}\mathcal{J}_{n}]$
	is the coupling constant for the contribution to dephasing of the
	sidebands beyond the RWA (we take the driving phase $\phi=0$ for
	simplicity). Assuming $\Omega_{\text{UV}}+m\omega\simeq\Omega_{\text{UV}}$
	and $\Omega_{\text{IR}}+m\omega\simeq m\Omega$, the integral for
	the symmetric part $\mathcal{S}\left(\Omega\right)$ of the correlation
	function is given by
	\begin{align}
		\frac{\Omega^{2}t^{2}}{2A}\int_{-\infty}^{\infty}d\Omega\mathcal{S}\left(\Omega-m\omega\right)\text{sinc}^{2}\left(\frac{\Omega t}{2}\right) & =\text{Ci}\left(\Omega_{\text{UV}}t\right)-\text{Ci}\left(m\omega t\right)-\cos\left(m\omega t\right)\text{Ci}\left(\Omega_{\text{UV}}t\right)+\sin\left(m\omega t\right)\text{Si}\left(\Omega_{\text{UV}}t\right)\nonumber \\
		& +\cos\left(m\omega t\right)\text{Ci}\left(\Omega_{\text{IR}}t\right)-\sin\left(m\omega t\right)\text{Si}\left(\Omega_{\text{IR}}t\right)+m\omega t\left[2\text{Si}\left(\Omega_{\text{UV}}t\right)-\frac{m\pi}{|m|}\right]\nonumber \\
		& -m\omega t\left[2\text{Si}\left(m\omega t\right)-\frac{m\pi}{|m|}\right]-\frac{1}{2}\left[1-\cos\left(m\omega t\right)\right]+\log\left(\gamma\right)-\log\left(\Omega_{\text{UV}}/m\Omega\right),\label{eq:big-integral-result}
	\end{align}
	where $\text{Si}\left(x\right)=\int_{0}^{x}dy\sin\left(y\right)/y$
	and $\text{Ci}\left(x\right)=-\int_{x}^{\infty}dy\cos\left(y\right)/y$
	are the sine and cosine integral functions, respectively. Assuming
	$t\Omega_{\text{IR}}\ll1$ and $t\Omega_{\text{UV}}\gg1$ we can employ
	the limit properties of these functions to reduce the problem. This
	yields the following expression for $f\left(t\right)$
	
	\begin{equation}
		f\left(t\right)\propto\exp\left\{ -\sum_{m\neq0}\frac{\partial F_{n,m}}{\partial\delta_{0}}\frac{\partial F_{n,\bar{m}}}{\partial\delta_{0}}\frac{A}{2\left(m\hbar\omega\right)^{2}}m\omega t\left[2\text{Si}\left(m\omega t\right)-\pi\text{sgn}\left(m\right)\right]\right\} ,\label{eq:decoherence-1}
	\end{equation}
	multiplied by a series of non-decaying oscillating factors that do
	not contribute to dephasing. Since $\text{Si}\left(x\right)\simeq\pi/2$
	for $x\to\infty$, after a few periods of the ac voltage this term
	reduces to zero and can be neglected in the high frequency approximation.
	The dephasing below this time scale is $\sim A\partial_{\delta_{0}}F_{n,m}\partial_{\delta_{0}}F_{n,\bar{m}}/(m\hbar\omega)^{2}$
	and is regardless negligible. To reach this result we employ\citep{Russ2015}
	$\Omega_{\text{UV}}t\gg1$ and $\Omega_{\text{IR}}t\ll1$, where $\Omega_{\text{UV}}$
	and $\Omega_{\text{IR}}$ are the ultraviolet and the infrared cutoffs,
	respectively, which are necessary to ensure convergence of the integral.
	Since we want to discuss also dephasing for longer times, we may reconsider
	the last condition. The next order contribution (in $\Omega_{\text{IR}}t)$
	has the form 
	\begin{equation}
		f\left(t\right)\propto\exp\left\{ -\sum_{m\neq0}\frac{\partial F_{n,m}}{\partial\delta_{0}}\frac{\partial F_{n,\bar{m}}}{\partial\delta_{0}}\frac{A\Omega_{\text{IR}}^{2}t^{2}}{8\left(m\hbar\omega\right)^{2}}\left[1-\cos\left(m\omega t\right)\right]\right\} ,\label{eq:decoherence-1-2}
	\end{equation}
	which exhibits the time periodic behavior that we expect from the
	oscillating terms and can be neglected for large frequencies. For
	$\Omega_{\text{IR}}t\gg1$ the $\text{Ci}\left(x\right)$ function
	that gives rise to this behavior vanishes. 
	
	For the contribution from the antisymmetric part of the correlation
	function $\mathcal{A}\left(\Omega\right)=\mathcal{S}\left(\Omega\right)\tanh\left(\hbar\Omega/2k_{B}T\right)$,
	we consider the classical limit $\hbar\Omega\ll2k_{B}T$ first. Then
	we can perform a series expansion of $\tanh\left(x\right)$ for $x\ll1$.
	The first order term is proportional to the integral
	
	\begin{equation}
		t^{2}\int_{-\infty}^{\infty}d\Omega\text{sinc}^{2}\left(\frac{\Omega t}{2}\right)\simeq\frac{2-2\cos\left(m\omega t\right)}{m\omega}-t\left[2\text{Si}\left(m\omega t\right)-\pi\text{sgn}\left(m\right)\right],
	\end{equation}
	which similarly vanishes after a number of ac periods. The third order
	term does not contribute to dephasing, either. On the other hand,
	in the $T\to0$ limit, the result is akin to Eq.~\ref{eq:decoherence-1}.
	Regarding the rest of the terms
	\begin{equation}
		\Delta\varphi^{\left(1\right)}\left(t\right)\simeq\sum_{m\neq m'\neq0}\frac{\partial F_{n,m}}{\partial\delta_{0}}\frac{\partial F_{n,m'}}{\partial\delta_{0}}\int_{-\infty}^{\infty}d\Omega\mathcal{S}\left(\Omega\right)\frac{-1}{\Omega-m'\omega}\frac{1}{\Omega+m\omega}\left[e^{-i\left(\Omega-m'\omega\right)t}-1\right]\left[e^{i\left(\Omega+m\omega\right)t}-1\right].
	\end{equation}
	Solving the integral results in a set of constant and oscillating
	terms which do not contribute to dephasing in the high frequency limit,
	in a similar way to Eqs.~\ref{eq:decoherence-1} and~\ref{eq:decoherence-1-2}. 
	
	The second order contribution $\Delta\varphi^{\left(2\right)}\left(t\right)$
	from the RWA Hamiltonian is given in Ref.~\citep{Russ2015} and has
	been included in the expression for the dephasing time in Eq.~\ref{eq:Deph-time}.
	The contribution from the neglected terms in Eqs.~\ref{eq:E-of-t}
	and~\ref{eq:J-of-t} due to the RWA approximation is given by
	
	\begin{align}
		\Delta\varphi^{\left(2\right)}\left(t\right) & \simeq\frac{1}{2}\sum_{m,m'\neq0}\frac{\partial^{2}F_{n,m}}{\partial\delta_{0}^{2}}\frac{\partial^{2}F_{n,m'}}{\partial\delta_{0}^{2}}\int_{0}^{t}dse^{im\omega s}\int_{0}^{t}ds'e^{im'\omega s'}\left\langle \hat{\xi}_{i}\left(s\right)\hat{\xi}_{i}\left(s'\right)\right\rangle ^{2}.\label{eq:accumul-phase-ac-2-1}
	\end{align}
	As for the first order, in the quasistatic limit the resulting contribution
	oscillates with the frequency of the voltage. As such, the main contributions
	to dephasing will be in the high frequency part of the correlation
	function. We focus on the $m=-m'$ terms. Then, the second order contribution
	is
	
	\begin{align}
		\Delta\varphi^{\left(2\right)}\left(t\right)= & \frac{A^{2}}{2\hbar^{2}\omega^{2}}\sum_{m\neq0}\frac{\partial^{2}F_{n,m}}{\partial\delta_{0}^{2}}\frac{\partial^{2}F_{n,\bar{m}}}{\partial\delta_{0}^{2}}I_{m}^{\left(2\right)}\left(t\right),\label{eq:accumul-phase-ac-2-2}\\
		I_{m}^{\left(2\right)}\left(t\right)= & \frac{\omega^{2}t^{2}}{A^{2}}\int_{-\infty}^{\infty}d\Omega\int_{-\infty}^{\infty}d\Omega'C\left(\Omega\right)C\left(\Omega'\right)\text{sinc}^{2}\left[\frac{\left(\Omega+\Omega'+m\omega\right)t}{2}\right].\label{eq:deph-2nd-order-int}
	\end{align}
	Similarly, the $m\neq-m'$ terms are
	
	\begin{align}
		\Delta\varphi^{\left(2\right)}\left(t\right)= & \frac{A^{2}}{2\hbar^{2}\omega^{2}}\sum_{m\neq m'\neq0}\frac{\partial^{2}F_{n,m}}{\partial\delta_{0}^{2}}\frac{\partial^{2}F_{n,m'}}{\partial\delta_{0}^{2}}K_{m,m'}^{\left(2\right)}\left(t\right),\label{eq:accumul-phase-ac-2-3}\\
		K_{m,m'}^{\left(2\right)}\left(t\right)= & \frac{-\omega^{2}}{A^{2}}\int_{-\infty}^{\infty}d\Omega\int_{-\infty}^{\infty}d\Omega'C\left(\Omega\right)C\left(\Omega'\right)\frac{\left[e^{-i\left(\Omega+\Omega'-m'\omega\right)t}-1\right]}{\Omega+\Omega'-m'\omega}\frac{\left[e^{i\left(\Omega+\Omega'+m\omega\right)t}-1\right]}{\Omega+\Omega'+m\omega}.\label{eq:deph-2nd-order-int-1}
	\end{align}
	
	\begin{figure}
		\begin{centering}
			\includegraphics[width=0.5\columnwidth]{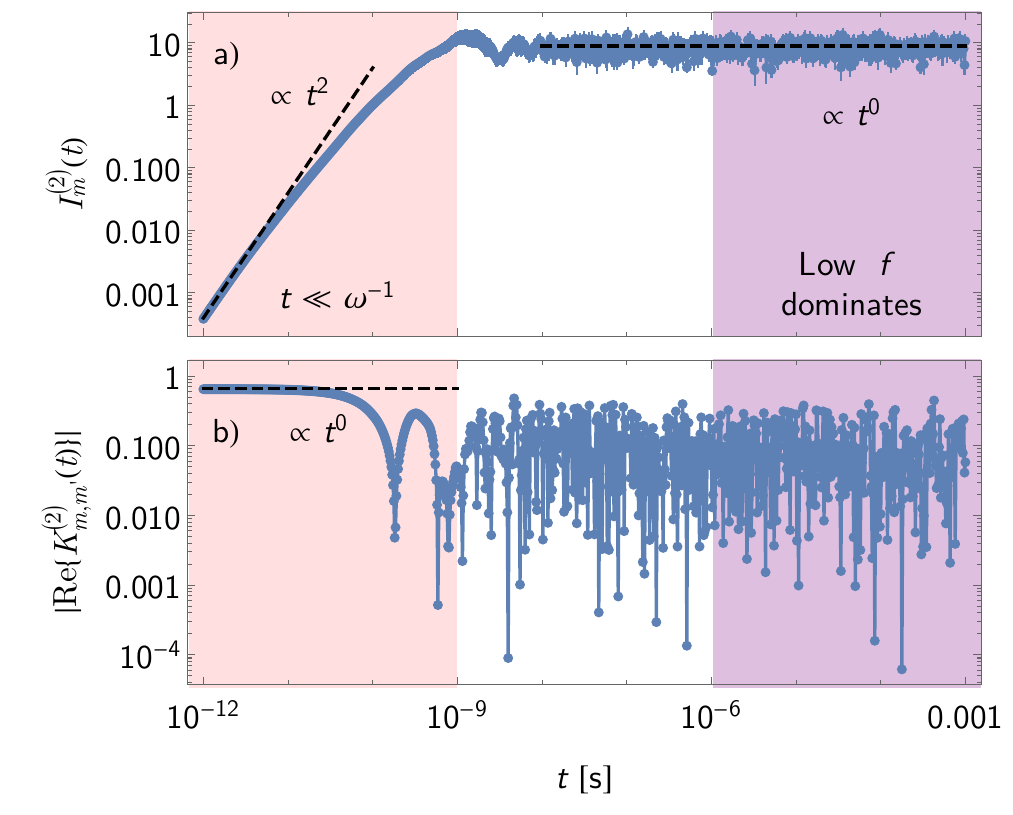}
			\par\end{centering}
		\caption{a) Integral $I_{m}^{\left(2\right)}\left(t\right)$ contributing to
			dephasing for $m=-m'$ (See Eq.~\ref{eq:deph-2nd-order-int}.) as
			a function of time for $T=40\,\text{mK}$, $m\omega=1\,\text{GHz}$,
			$\Omega_{\text{IR}}=10^{3}\,\text{Hz}$ and $\Omega_{\text{UV}}=1\,\text{THz}$
			(error bars reflect integration errors in the quasi-Monte Carlo approach
			employed to calculate them). b) Integral $K_{m,m'}^{\left(2\right)}\left(t\right)$
			contributing to the dephasing time for $m\protect\neq-m'$ (See Eq.~\ref{eq:deph-2nd-order-int-1}.)
			for the same parameters as $a$, with $m'=3m$. \label{fig:dephasing-app}}
	\end{figure}
	
	For the estimation of these integrals we resort to numerical integration.
	We have represented the integral $I_{m}^{\left(2\right)}\left(t\right)$
	in Fig.~\ref{fig:dephasing-app}~a). The behavior of the integral
	can be separated in a region $t\ll\omega^{-1}$ ($m\omega=1\,\text{GHz}$
	in this case) where the behavior is similar to the quasistatic case,
	with a limiting dependence $\propto-t^{2}$ as in Eq.~\ref{eq:accumul-phase}
	(represented with a dashed black line in the figure), an intermediate
	regime with $t\sim\omega^{-1}$ where the integral oscillates and
	a long time behavior as $t\gg\omega^{-1}$ which is dominated by the
	low frequency behavior of the integral and does not contribute to
	dephasing. Hence, as discussed for Eq.~\ref{eq:decoherence-1}, we
	expect that in the high-frequency regime the dephasing due to these
	second-order terms can also be neglected for time scales larger than
	the ac bias. We have also represented the integral $K_{m,m'}^{\left(2\right)}\left(t\right)$
	in Fig.~\ref{fig:dephasing-app}~b) for $m'=3m$. In this case,
	most of the contribution is  constant in the regime $t\ll\omega^{-1}$.
	We remark that these contributions to the accumulated phase include
	both the effect of noise on the time evolution during the micromotion
	and the higher order corrections to the RWA Hamiltonian (e.g. Eq.~\ref{eq:high-freq-expansion}). 
	
\end{widetext}
~
\end{document}